# Angular Momentum of Binary Asteroids:

# Implications for their possible origin

P. Descamps[a], F. Marchis[a,b]

[a] Institut de Mécanique Céleste et de Calcul des Éphémérides, Observatoire de Paris,
77 avenue Denfert-Rochereau, F-75014 Paris, France
[b] University of California at Berkeley, Department of Astronomy, 601 Campbell Hall, Berkeley,
CA 94720, USA

Pages: 36

Table: 1

Figures: 3

Editorial correspondence to:

Pascal Descamps

IMCCE – Paris Observatory

77, avenue Denfert-Rochereau

75014 Paris

France

Email: descamps@imcce.fr

# ABSTRACT

We describe in this work a thorough study of the physical and orbital characteristics of extensively observed main-belt and Trojan binaries, mainly taken from the LAOSA (Large Adaptive Optics Survey of Asteroids, Marchis et al., 2006c) database, along with a selection of bifurcated objects. Dimensionless quantities, such as the specific angular momentum and the scaled primary spin rate, are computed and discussed for each system. They suggest that these asteroidal systems might be the outcome of rotational fission or mass shedding of a parent body presumably subjected to an external torque.

One of the most striking features of separated binaries composed of a large primary ($R_p > 100$ km) with a much smaller secondary ($R_s < 20$ km) is that they all have  total angular momentum of ~0.27. This value is quite close to the Maclaurin-Jacobi bifurcation (0.308) of a spinning fluid body. Alternatively, contact binaries and tidally locked double asteroids, made of components of similar size, have an angular momentum larger than 0.48. They compare successfully with the fission equilibrium sequence of a rotating fluid mass. In conclusion, we find that total angular momentum is a useful proxy to assess the internal structure of such systems.

# Keywords

## 1. Introduction

During the last decade more than one hundred binaries have been discovered among near-Earth asteroids (Pravec et al., 2006, Richardson and Walsh, 2006), main-belt asteroids (Merline et al., 2002), Kuiper-belt objects (Noll, 2006) and through a wide variety of techniques, including adaptive optics (AO), direct imaging, photometric lightcurve and radar astronomy among others. These gravitationally bound binaries provide new opportunities to study the internal structure and the formation scenarios of minor planets (small solar system bodies) in connection with the origin of our Solar system. However, mostly main-belt binaries have been fully and accurately characterized with respect to their orbits and physical parameters (system mass, size, density, primary shape) being well constrained (e.g. Marchis et al., 2003, Marchis et al., 2005a, Marchis et al., 2005b, Descamps et al., 2007a, Marchis et al., 2007). Their orbital parameters and physical characteristics published so far have tremendous variety with orbital periods ranging from 0.5 to 80 days, secondary-to-primary size ratios from 0.1 to 1, density ranging from 0.7 to 3.5 g/cm³. Nevertheless this diversity also hides dynamical similarities between these systems. Indeed, all of these orbits are prograde with small or negligible inclination and eccentricity.

In this present work, we will attempt to draw conclusions about trends among these binary objects through the use of comparable dimensionless parameters, scaled spin rate and specific angular momentum, with the intent of shedding light on the origin, evolution and internal structure of binary asteroid systems. The specific angular momentum is one of the most fundamental dynamical parameters, witness in whole or in part to the past dynamical history. We chose to restrict our study to a selected set of main-belt and Trojan asteroids listed in Section 2 for which their main physical properties (sidereal spin period, size, density and shape) are known with an accuracy to the 10% level.

Sections 3 and 4 propose a dychotomic analysis according to the amount of total specific angular momentum with which binary asteroids are endued. We interpret these data and discuss the origin of the binary systems with a small satellite as well as the puzzling bifurcated objects and similarly-sized systems in the framework of the stability theory of idealized ellipsoidal equilibrium figures with uniform internal density.

## 2. Selected binary asteroids

### 2.1 The Large Adaptive Optics Survey of Asteroids (LAOSA)

LAOSA is designed to study minor planets using high angular resolution capabilities provided by large aperture telescopes with AO systems available on the Very Large Telescope (VLT-8m UT4) and the Gemini North 8m telescope (Marchis et al., 2006c). Ten of the most interesting targets of this extensive database, containing ~1100 observations, are binary Main-Belt Asteroids (MBAs). Most of these LAOSA binary systems have reliable orbit determination and spin pole solution. The binary systems have prograde orbits (orbit pole aligned with spin axis) with small eccentricity and inclination. Because of the intrinsic limits of the AO technology, the asteroids observed were large ($D_p \geq 90$km). Those found to be binary had widely disparate components with brightness variations up to 8.5 magnitudes fainter than the primary ($D_s \geq 0.02D_p$) and separated by at most 10 primary radii. Since their bulk density (ranging from 0.7 to 3.5 g/cm$^3$) is generally lower than the one measured for their meteorite analogs, they should have a high porosity. The binary minor planets of the LAOSA database for which we have accurate orbit solutions are (22) Kalliope, (45) Eugenia, (87) Sylvia, (90) Antiope, (107) Camilla, (121) Hermione, (130) Elektra, (283) Emma, (379) Huenna and (617) Patroclus (references included in Table 1). Two of these binary systems, (45) Eugenia (Marchis et al., 2007c) and (87) Sylvia (Marchis et al., 2005a) are known to have additional components. To simplify this work without loss of accuracy, we did not take into consideration the tertiary component in our analyses.

## 2.2 Non LAOSA bodies incorporated in the study

In addition to the LAOSA asteroids, we include 9 other objects in our study: (243) Ida, (216) Kleopatra, (624) Hektor, (3169) Ostro, (69230) Hermes, (854) Frostia, (1089) Tama, (1313) Berna, and (4492) Debussy.

In 1993, the Galileo spacecraft discovered the first binary asteroid, (243) Ida, a 30km diameter S-type asteroid with a small satellite named Dactyl (Belton et al., 1995, Chapman et al. 1995). It is unclear whether or not Ida should be considered as a rubble pile asteroid (Britt et al. 2002), meaning that its internal structure is defined as a weak aggregate of diverse size components held together by gravity rather than material strength. We adopt a density of $2.4 \pm 0.1$ g/cm$^3$ following the conclusions of Petit et al. (1997) based on orbit long-term stability requirements.

(216) Kleopatra is a M-type asteroid considered as a rubble-pile. The first image of (216) Kleopatra using a Maximum Entropy algorithm was reported by Storrs et al. (2000), who found it to be elongated by at least 4:1 (projected on the sky at the time of observation). AO observations revealed the double-lobed, narrow-waisted shape of Kleopatra (Marchis et al., 2000). Radar observations helped to develop a 3D-shape model for this system (Ostro et al., 2000) putting a constraint on its surface bulk density no less than 3.5g/cm³, indicating that Kleopatra could be the outcome of collisional events (Hestroffer et al., 2002, Tanga et al., 2001), and much of its interior might have an unconsolidated rubble-pile structure (Ostro et al., 2000). We adopt such a value with an arbitrary uncertainty of 1 g/cm³ as the internal bulk density, knowing that it could be a lower limit.

The Trojan asteroid (624) Hektor was the first object suspected to be a contact binary from its large amplitude lightcurve in early observations (Dunlap and Gehrels, 1969, Cook 1971, Hartmann and Cruikshank, 1978, Weidenschilling, 1980). It was only recently that Marchis et al., (2006b) confirmed this hypothesis by directly imaging the minor planet and revealing its typical bilobated shape. Simultaneously, these authors release the discovery of a small satellite orbiting Hektor which permitted a preliminary estimation of its bulk density of 2.2±0.5g/cm³, inferred from Kepler's third law. This value is in fairly good agreement with the one inferred by Weidenschilling (1980) by modeling the observed rotational lightcurves as a contact binary of two Darwin ellipsoids.

Another contact binary included in this study is the 11km-diameter main belt asteroid (3169) Ostro (Descamps et al., 2007b). Gathering photometric observations taken in 1986, 1988 and 2005, they derived a simple model using Roche ellipsoids with an inferred bulk density of 2.6±0.4 g/cm³.

(69230) Hermes is a potentially hazardous asteroid (PHA). It is the only binary near-Earth asteroid (NEA) known to have reached a fully synchronous state in which the spin periods of the components are equal to the revolution period of the components about their barycentre (Margot et al., 2003, Pravec et al., 2003). The low  amplitude lightcurve of 0.06 mag (Pravec et al., 2006) implies that the two components seen by radar are nearly spherical despite their small sizes (630 and 560m).

The last objects incorporated in the dataset are the four small synchronous binary systems (854) Frostia, (1089) Tama, (1313) Berna, and (4492) Debussy reported by Behrend et al. (2006). However, the authors confused the semi-major axis with the semi-separation[1]. Therefore in the present work we correct this error applying a factor of two to the reported separations. Other such systems have been discovered recently but, at the time of writing, their parameters remain incomplete (Kryszczynska et al., 2005, Manzini et al., 2006).

## 2.3 Total angular momentum

In appendix A, it is shown that the specific angular momentum of a binary system, made of a primary body presumed to be spinning around the principal axis of maximum inertia and a secondary body orbiting in the equatorial plane, is given by the following relation (cf. appendix A.3):

$$H = \frac{q}{(1+q)^{\frac{13}{6}}} \sqrt{\frac{a(1-e^2)}{R_p}} + \frac{2}{5} \frac{\lambda_p}{(1+q)^{\frac{5}{3}}} \Omega + \frac{2}{5} \lambda_s \frac{q^{\frac{5}{3}}}{(1+q)^{\frac{7}{6}}} (\frac{R_p}{a})^{\frac{3}{2}}$$

---

1   The error on the semi-major axis does not influence our interpretation of the density of an object since the two errors cancel out  in  equation (5) of Behrend et al. (2006). The corrected expression is :

$$a = \frac{r_b}{2 \sin(\Delta \Phi_g / 4)}$$

where $r_b$ is the smallest semi-major axis of each prolate ellipsoid and $\Delta\Phi_g$ is the relative duration of an event expressed in angular form.

Where $q$ is the secondary-to-primary mass ratio, $a$ the semimajor axis, $R_p$ the primary radius and $e$ the eccentricity. The normalized spin rate is defined by the ratio $\Omega = \omega_p/\omega_c$, where $\omega_p$ is the primary rotation rate and $\omega_c$ the critical spin rate for a spherical body (cf. Appendix A). The effect of an asteroid's nonsphericity, which in general will increase the angular momentum corresponding to a specific spin period, is given by the parameter $\lambda$ which is the ratio between the moment of inertia of the body with respect to its spin axis and the moment of inertia of the equivalent sphere (see appendix A.2). For convenience the secondary is considered as spherical so that $\lambda_s = 1$. For primaries represented by triaxial ellipsoids of uniform mass density with semi-axes $a_0 > a_1 > a_2$, which are the asteroid principal axes, we have:

$$\lambda = \frac{1 + \beta^2}{2(\alpha\beta)^{2/3}}$$

where we define the axis ratios $\alpha = a_2/a_0$ and $\beta = a_1/a_0$ so that $\alpha \leq \beta \leq 1$.

For each object, all physical parameters and dimensionless quantities $\lambda_p$, $\Omega$ and $H$ are listed in Table 1. Unless otherwise stated, the nonsphericity parameter $\lambda_p$ is derived from the ellipsoidal shape models, provided by the database maintained at the Poznan Observatory by Agnieszka Kryszczynska[2]. From lightcurve inversion technique (Kaasalainen et al., 2001), some of the selected asteroids have been portrayed by means of a topographic shape solution through a polyhedron surface (Kaasalainen et al., 2002, Marchis et al., 2006d, Michalowski et al., 2006), the mass properties of which, the inertia tensor and the nonsphericity parameter, have been computed assuming a homogeneous mass distribution.

---

[2] Table and related documentation are available from http://www.astro.amu.edu.pl/Science/Asteroids.

Following Weidenschilling (1981), we plot in Fig. 1 the scaled spin rate $\Omega$ versus the total specific angular momentum $H$ with hopes of gaining insight for a more global interpretation of how these quantities relate. The equilibrium sequences of rigidly rotating fluids are represented in the same diagram. Among those depicted are, on one hand, the Maclaurin spheroids and the Jacobi ellipsoids for single bodies and, on the other hand, the Roche and Darwin ellipsoids[3] which are tidally distorted figures associated with doubly synchronous rotation (see Chandrasekhar, 1969 for a summary of much of this subject as of 1968). In view of the close vicinity of selected asteroids with equilibrium sequences, such families of idealized bodies should play a role in elucidating the origin and evolution of observed and well-defined binary systems.

Materials:
**[INSERT here Table1]**
**[INSERT here Fig.1]**

### 3. High size ratio binary asteroids

### 3.1 Outstanding facts

.

Taking a close look at the data listed in the Table 1 pertaining to the binary main belt asteroids of the LAOSA database, we notice that despite the diversity of their physical and orbital characteristics, they exhibit a similar specific angular momentum estimated to H~0.27 with a non-sphericity parameter of λ~1.25 and an adimensional spin rate $\Omega$ between 0.50 and 0.56. These features are so remarkable as to be worthy of attention and compared with other analysis.

---

3 Darwin's problem differs from Roche's problem in a sense that it seeks to allow for mutual tidal interactions between both components while Roche's problem is only concerned with the tidal effect of a solid sphere. For the Roche ellipsoids, one can construct sequences characterized by some fixed value of the mass ratio *s*. For the Darwin ellipsoids, at first sight, it appears that only two ratios of masses are possible, 0 or 1.

Pravec et al. (2006) pointed out that small asynchronous binary NEAs had a specific angular momentum close to 0.4, the disruption critical limit for a single spherical body. Therefore we conclude in the first place that large binary main-belt asteroids possibly have a different nature and/or origin than these small binary NEAs.

From the Figure 1, we notice that these LAOSA objects concentrate in a region close to the Maclaurin-Jacobi bifurcation which occurs at H=0.308. This bifurcation is defined as the largest angular momentum for which a fluid ellipsoid is stable when it is a figure of revolution[4]. Theoretically, a secular instability sets in this region stimulated by the presence of viscosity in a fluid should result in a slow departure from the unperturbed figure leading to the Jacobi family. These concepts are summarized by Farinella (1981) and have been used repeatedly in the search of Jacobi equilibrium shapes among binary asteroids (Hestroffer, 2004, Hestroffer and Tanga, 2005). This search was stimulated by the idea that most asteroids are the outcome of a gravitational reaccumulation process after the fragmention of a parent body through cataclysmic collisions (Davis et al., 1979). In this scenario, the primary of these binary asteroids is described as a collection of boulders, only held together by gravity (Chapman, 1978), reminiscent of a fluid body. To date, no asteroids have been satisfactorily identified with figures closely related to Jacobi ellipsoids. The present work based on a large number of well-studied orbits of binary asteroids supports this conclusion. Alternatively, Pravec and Harris, (2000) claim that these asteroids should not be in hydrostatic equilibrium due to internal friction. Our study shows, however, that these binary asteroids are accumulated in a region close to the Maclaurin-Jacobi transition suggesting that their behaviour is somehow linked with hydrostatic equilibrium. The renovated approach of the internal structure of a cohesionless body proposed by Holsapple (2001, 2004) and discussed in section 3.2, may help to interpret and understand the observed behaviour.

---

4 Axially symmetric figure where $a_0 = a_1$ Body with such a shape is also called an oblate spheroid.

### 3.2 The Mohr-Coulomb model of rubble-pile asteroids and stability

Holsapple (2001) derived the limits on equilibrium configurations for a cohesionless, but not strengthless, solid ellipsoidal rotating body. The body is modeled with elastic-plastic materials governed by a Mohr-Coulomb yield[5] model characterized by its angle of friction. This model is appropriate for "rubble-pile" asteroids that consist of a continuum assemblage of particles held together at moderate densities by mutual gravitational forces. As a consequence, Holsapple derived lower spin limits than the value given by the simple comparison of gravity and spin forces at the equator, where the failure of a body is caused by shearing instead of tensile forces. In other words, a body will fail in shear before it can reach the "rotation rate barrier" of 2.1h/rev derived by Harris (1996) for an assumed typical density of $2.5g/cm^3$ and before material can be flung off the equator. In 2004, Holsapple derived new criteria for the loss of stability of an equilibrium state thanks to a new stability approach based on energy considerations. A body can become dynamically unstable if it has some strength but it is not clear whether it will be disrupted or reach a new equilibrium state by a large change of configuration. Figures 2 and 3 reproduce the main results derived by Holsapple for oblate ($\beta=1$) and prolate ($\alpha=\beta$) asteroids where we plot only the LAOSA binaries discussed in this section. In the case of oblate bodies (Fig.2), for states above the Maclaurin curve, (e.g with a non-zero friction angle $\phi$), the spin is greater than that allowable in a fluid body at that same shape. Any state between limit curves for a given friction angle is possible without yield. When yielding occurs, it leads to an increased oblateness and, in the diagram, the body will move to the left along a line of constant angular momentum. In addition, with this deformation, there is a volume expansion and consequently a lowering of the bulk density for all angles of friction. The case of prolate bodies (Fig. 3) is not qualitatively different.

---

5    The yield point is defined as the stress at which a material begins to plastically deform. Once the yield point is passed some fraction of the deformation will be permanent and non-reversible.

Materials:
**[INSERT Fig.2]**
**[INSERT Fig.3]**

Asynchronous binary LAOSA objects have intermediate shapes between oblate and prolate figures. They have been superimposed to the limit curves in figures 2 and 3. In each figure they conspicuously accumulate just below the stability limit along a line of constant angular momentum originating from a point on the curve for which the minimum friction angle is about $\varphi \sim 11\text{-}14°$ (intersection between the iso-angular momentum curve H~0.27 and the stability limit). This result is unexpected because typical terrestrial granular materials have $\varphi \sim 35°$ (Strahler, 1971). Bodies with such a friction angle are allowed to occupy all states between the lower and upper corresponding limit curves. Consequently, we would expect to find binary asteroid states covering a larger range of angular momentum values.

A plausible evolutionary scenario might involve successive spin-up, driving a body up to its equilibrium limit curve where deformation will occur, giving rise to a more flattened and less dense state. It could then gradually reach the stability limit at a point where the corresponding limit curve will intersect. It is natural, therefore, to expect that the instability is capable of extracting material into orbit and ultimately transforming it into a binary system. Weidenschilling et al. (1989) and Merline et al. (2002) already pointed out that satellites formed by orbiting ejecta would be small (size ratio < 0.2) in prograde orbits with respect to the primary's spin about rapidly spinning primaries. The only requirements are that the primary spin period should be smaller than ~6h and that its shape be significantly non-spherical which is corroborated by the observational facts. However, regarding their angular momenta and the level of non-sphericity no constraints were inferred. To date, the mechanism of yielding small satellites by accretion of ejecta after an impact fails to be numerically reproduced (Durda and Geissler, 1996). The similar global properties revealed by the present work, unidentified so far, substantiate such a mechanism and give new indication on the important role played by an appropriate internal strength model of involved bodies.

## 4. Synchronous double asteroids

### 4.1 Overview

The objects listed in this section may be divided into two parts. The first part consists of strictly speaking synchronous double asteroids which are binary systems with nearly equisized components maintaining the same face towards each other at all times. The second part is devoted to the so-called contact binaries which may be considered as a peculiar case of double asteroids with components so near to one another that they overlap to form a single body.

**Synchronous binaries**

After the discovery in the Main belt of the first doubly synchronous system, (90) Antiope (Merline et al. 2000), it was then thought that this puzzling system might be unique in the Solar system regarding its large size, relatively young age, spin-locked state and the nearly perfect equilibrium shape of its components (Weidenschilling et al. 2001). However, soon afterwards, another double system with the same global characteristics was discovered among a different asteroidal group, the Trojan (617) Patroclus (Merline et al., 2001). In 2003, thanks to Arecibo and Goldstone radar data, an additional fully synchronized system was brought out with the small 600m in diameter NEA (69230) Hermes (Margot et al., 2003, Pravec et al., 2003). To date, apart from the Pluto-Charon system, no new synchronous binary transneptunian objects (TNOs) have been detected. AO imaging of objects of equal brightness on the world's largest 8-10m groundbased telescopes achieves an angular separation as small as 0.07 arcsec. This is precisely the angle subtended by two objects 50 km apart observed at a distance of 1 AU which rules out most of the NEAs and TNOs assuming a separation of at most 10 times either mean radius.

Nevertheless, another powerful technique to reveal synchronous binarity is the photometric observation of mutual events. Such events within a double system may occur provided that it is viewed from a nearly edge-on configuration. If the system is fully synchronized, for which rotational periods are akin to the orbital period, these events will yield a single-period lightcurve[6]. This technique is well suited for small asteroids and requires only moderate aperture telescopes. Such photometric detections were recently performed providing six new synchronous systems (Kryszczynska et al., 2005, Behrend et al., 2006, Manzini et al., 2006). Furthermore, we can take advantage of a long-term AO astrometric monitoring of large double asteroids to predict occurrence of mutual events in order to very accurately characterize some of their physical properties required for the study at hand (Descamps et al. 2007a, Marchis et al., 2007a). In the Figure 1 synchronous systems (labeled "Roche systems"), amazingly spread along the same equilibrium sequence of two fluids approximated by two identical ellipsoids rotating around their center of gravity. This family of idealized figures of simple binary systems

---

6 A lightcurve which contains only one period produced by the equality between orbital period and spin periods.

have been addressed and detailed by Roche in 1849 and Darwin in 1887.

**Contact binaries**

The second remarkable group of interest is composed of contact binaries. Several Kuiper-belt objects (Sheppard and Jewitt, 2004), MBAs (Descamps et al., 2007b) and NEAs (Polishook and Brosch, 2006) have also revealed lightcurve morphology typical of candidate contact binaries viewed edge-on. Among them, 2002 NY 40 (Howell et al., 2003) and 2005 CR37 (Benner et al., 2006) have been conclusively confirmed by radar imagery to look like peanut-shaped single bodies. Three of these objects, (216) Kleopatra, (624) Hektor and (3169) Ostro, are in our sample and are found in Figure 1 to lie in a narrow region centered on the point of angular velocity of 0.33 with a specific angular momentum of about 0.48.

### 4.2 Equilibrium figures

In stellar astrophysics, the fission theory has been developed and used to explain the high proportion of binary stars as the dynamical consequence of the contraction of a rotating, self-gravitating fluid mass. The concept of the model is relatively simple: a single mass star becomes elongated and then divides into a pair of detached masses due to its spin. In the present work, this concept is revitalized in the context of binary asteroids.

In the 1980s, from numerical investigations, Eriguchi et al. (1982) discovered that, along the Jacobi sequence, the fourth harmonic instability ($\Omega = 0.3995$; $H = 0.48412$) smoothly led to the bifurcation from Jacobi ellipsoids to a dumbbell-shaped sequence (see point A in Fig. 1). In other words an elongated fluid body will evolve toward a peanut shape with two equal lobes at the extremities and a furrow, a "bridge" of material, in the middle. Farther along the sequence, the furrow deepens to shrink to nothing and a binary state is realized. They calculated the binary sequence of two equal-mass detached objects as well as binary sequences with different mass ratios of the components (Hachisu and Eriguchi, 1984). The equal-mass binary sequence (Darwin with $q$=1) merges with the dumbbell sequence. The point of contact (see point B in Fig. 1) of the two sequences represents an unstable equilibrium (contact binary). The two branches ultimately form one continuous sequence, the so-called fission sequence. This new equilibrium sequence along with Roche sequences at various $q$, obtained for different size ratios, are plotted in Fig. 1. It is of interest to point out that no synchronous double systems with unequal components, but with a size ratio larger than 0.5, have been observed so far. Pravec et al. (2006) already noted that asynchronous binary NEA systems concentrate below $D_s/D_p = 0.5$. Weidenschilling et al. (1989) pointed out that only systems with large satellites (size ratio ~ 1) can be tidally stable and maintain spin-orbit synchronism. Accordingly, nothing prevents synchronous systems with size ratio of 0.7, 0.6 or even 0.5 from existing. One possible explanation to account for the lack of such systems is to consider that double synchronous asteroids are made of two detached congruent pieces originated from one single parent body according to the aforementioned fission process.

The distribution of the contact binaries and double systems strikingly matches the fission sequence (Fig.1). One can show that for all the bodies with scaled spin rate <~0.2, the orbital angular momentum dominates over the spin and the lower region of that curve is indistinguishable from any pair of orbiting spheres (see appendix A). However, recent extensive work on the system of (90) Antiope showed that Roche formalism is not only a convenient way of modeling a binary synchronous system but also provides information about the object's internal structure (Descamps et al., 2007a). The bulk density, directly measured from the orbit analysis, was in excellent agreement with the density inferred from the Roche hypothesis and the figures of the components were adequately matched by the corresponding Roche ellipsoids. Furthermore, if we refer to the other known double synchronous asteroids, although their shapes stay uncertain, their observed lightcurves with low amplitudes, 0.06mag for Hermes (Pravec et al., 2006) and 0.09mag for Patroclus (Mottola, 1991, personal communication), suggest that either components of these bodies are nearly spherical, just slightly squashed into ellipsoids with low equatorial elongation, in good agreement with the Roche models (Descamps, 2007c). Preliminary analysis of collected lightcurves in the framework of the 2007 international observing campaign[7] of mutual events within the system of (617) Patroclus (Marchis et al., 2007a), taking advantage of an edge-on aspect, supports that conclusion. Likewise, the same statement can be made about the contact-binaries (624) Hektor (Weidenschilling, 1980) and (3169) Ostro (Descamps et al., 2007b) the shapes of which may be matched by equilibrium ellipsoids. Actually, the departure of the real bodies' figures from the ellipsoids of the involved Roche problem chiefly depends on the internal friction level (Descamps, 2007c).

In summary, we state that a possible affiliation between contact-binaries and double asteroids, considered in the present work, may exist, endorsed by nearly perfect equilibrium shapes with size ratios very close to unity.

---

[7] Predictions and lightcurves are available on the devoted web page devoted to the 2007 mutual events on (617) Patroclus: http://www.imcce.fr/page.php?nav=en/observateur/campagnes_obs/patroclus/

## 5. Conclusion

Through the computation of the total angular momentum of some binary asteroids, a particular emphasis is placed on common features which imply that those systems share a common dynamical origin. Indeed, a striking behavioural analogy with rotating fluid mass properties is highlighted.

The binary asteroids selected in this study are thought to be weak gravitational aggregates or at least to have low density. We postulate that the cluster of binaries with significantly non-similar sized components, located near the bifurcation point, are examples of bodies governed by the theory of a fluid, self-gravitating, rotating ellipsoid modified by the Mohr-Coulomb failure criteria for cohesionless granular materials (Holsapple, 2001, 2004). Such bodies have stability limits depending on their shapes and spins which, once reached, might lead to a mass shedding process which could originate in satellite formation.

Our study shows that the contact binaries and synchronous double asteroids could be members of a similar evolutionary scenario, mimicking the rotational fission of a rigidly spinning fluid mass. However, a major issue which should be addressed which respect to the fission scenario is the requirement for an asteroid to surpass the third-harmonics instability (*H=0.39*), only valid for perfectly fluid bodies (Cartan, 1928), in order to arrive to the fourth-harmonics point of bifurcation from which it could undergo fission (*H=0.48*). We propose in the future to investigate the occurrence of the third-harmonics instability in the case of low internal friction bodies assuming the Mohr-Coulomb failure criteria.

Is there a connection between those two groups of binaries ? In the context of such a fission theory, it is implicitly assumed that such loosely consolidated objects evolve along the Jacobi ellipsoids family however, no asteroids have been identified along this line of evolution. Only the small asteroid (243) Ida appears to be on the Jacobi sequence (Fig.1) suggesting that it could be purely coincidental. In the framework of the Mohr-Coulomb strength model, however, the shape of a cohesionless body is no constrained by the Jacobi figure and a variety of shapes are permissible. Furthermore, among the contact binary population, at least one of them[8], (624) Hektor (Marchis et al., 2006), has been recently observed with a moonlet. One interesting possibility is that they are the outcome of a preceding mass-shedding instability encountered near the bifurcation point, meaning that these systems could be the "missing link" between the two groups described in our study. The source of increase for the total angular momentum remains a mystery. Tidal effects during planetary flybys (Walsh and Richardson, 2006) and the YORP effect (Cuk et al., 2007) are often invoked to explain the change of angular momentum of near-Earth objects and/or small main-belt asteroids (D<30 km). It is obviously beyond the scope of this paper to propose a complete explanation about the source of increase for the angular momentum.

Our study based on the angular momentum of these binaries shows a clear dichotomy between two groups of binary asteroids. Continuing observations to determine more orbital and physical parameters of all known binary systems (135 are known at the time of writing), including the near-Earth binaries and the Kuiper-belt binaries, and numerical investigations must certainly be pursued to gain insight into how these binary minor planets formed.

---

8 Another one could be the binary asteroid (121) Hermione the primary of which is suspected to be bilobated (Marchis et al., 2005b)

**Acknowledgements**

This work was partially supported by the National Science Foundation Science and Technology Center for Adaptive Optics, and managed by the University of California at Santa Cruz under cooperative agreement No. AST-9876783. This material is partly based upon work supported by the national Aeronautics and Space Administration issue through the Science Mission Directorate Research and Analysis Programs number 070.C-0746, 071.C-0669, 072.C-0753, 073.C-0851, 070.C-0458, 072.C-0016, 073.C-0062, 273.C-5034, 074.C-0052, 077.C-0422. We are grateful to S. Weidenschilling for reviewing this manuscript and to S. Kern for her exceptional work as a reviewer which improved significantly the quality of this article.

**Appendix A: Derivation of the total specific angular momentum of a binary system**

It will be convenient to refer to the total mass ($M$) of a binary system as the sum of the mass of the two components with an equivalent radius $R_e$. Then assuming the same bulk density $\rho$ for each body

$$R_e^3 = R_p^3 + R_s^3 \ (1)$$

and the total mass of the system reads

$$M = \frac{4}{3} \pi R_e^3 \rho = (1 + q) M_p \quad (2)$$

with $M_p$ the mass of the primary and $q$ the secondary to primary mass ratio, $q = (R_s/R_p)^3$.

The specific angular momentum is given by reference to the angular velocity of a sphere of the same mass as the sum of our two masses, spinning so as to have the same angular momentum. By this notation the angular momentum and the angular velocity are comparable between multiple systems and equilibrium figures of a single body. It is a dimensionless quantity obtained by dividing by the following term, notation first introduced by Darwin (1887):

$$\sqrt{GM^3 R_e} = MR_e^3 \sqrt{\frac{4}{3} \pi \rho G} = MR_e^3 \ \omega_c \ (3)$$

where $G$ is the gravitational constant and $\rho$ the bulk density. $\omega_c$ is the surface orbit frequency for a rigid spherical body, or the maximum spin rate that can be sustained by an undeformable body. At this spin rate, centrifugal forces would equal gravity at the equator of a spherical body.

Let $H_p$, $H_s$ and $H_o$ be the specific angular momenta of the primary, the secondary and the orbit respectively. If we assume that a satellite orbits its primary equatorially  in a prograde manner, the total angular momentum is then $H_t = H_p + H_s + H_o$.

## A.1 Orbital momentum

The whole system revolves orbitally about the centre of inertia described by a Keplerian elliptical orbit of eccentricity $e$ with an angular velocity $\omega_o$. Hence the orbital momentum is

$$H_o = \frac{4}{3}\pi\sqrt{1-e^2}[\,\omega_o R_p^3 d^2 + \omega_o R_s^3 D^2\,] \quad (4)$$

with

$$d = \frac{R_p^3 a}{R_p^3 + R_s^3}, D = \frac{R_s^3 a}{R_p^3 + R_s^3} \quad (5)$$

Hence, equation (4) becomes

$$H_o = \frac{4}{3}\pi\sqrt{1-e^2}\,\frac{R_p^3 R_s^3}{R_p^3 + R_s^3}\omega_o a^2 \quad (6)$$

which simplifies to

$$H_o = \frac{4}{3}\pi\,\frac{R_p^3 R_s^3}{R_p^3 + R_s^3}\cdot\frac{\omega_c}{\omega_o a^2 MR_e^3}\sqrt{1-e^2} \quad (7)$$

according to Kepler's third law

$$\omega_o = \frac{\sqrt{GM}}{a^{\frac{3}{2}}} = (\frac{R_e}{a})^{\frac{3}{2}}\omega_c = (\frac{R_p}{a})^{\frac{3}{2}}(1+q)^{\frac{1}{2}} \quad (8)$$

Finally, the specific orbital angular momentum reads

- 23 -

$$H_o = \frac{q}{(1+q)^{\frac{13}{6}}} \sqrt{\frac{a(1-e^2)}{R_p}} \quad (9)$$

Now let us consider the rotational angular momenta of each body.

## A.2 Rotational angular momentum of each body

We introduce the non-sphericity parameter $\lambda$ defined as the ratio between the moment of inertia of the body with respect to its spin axis, considered as the maximum moment of inertia axis, and the moment of inertia of the equivalent sphere by:

$$\lambda = \frac{kMR_e^2}{\frac{2}{5}MR_e^2} = \frac{5}{2}k \quad (10)$$

For the primary body we have

$$H_p = \frac{\frac{2}{5}\lambda_p M_p R_p^2 \omega_p}{MR_e^2 \omega_c} = \frac{2}{5}\frac{\lambda_p}{(1+q)^{\frac{5}{3}}}\frac{\omega_p}{\omega_c} \quad (11)$$

and for the secondary body

$$H_s = \frac{\frac{2}{5}\lambda_s M_s R_s^2 \omega_s}{MR_e^3 \omega_c} = \frac{2}{5}\lambda_s \left(\frac{R_s}{R_e}\right)^2 \frac{q}{(1+q)}\frac{\omega_o}{\omega_c} \quad (12)$$

Assuming a tidal spin-orbit synchronization of the secondary, i.e. $\omega_o = \omega_s$

$$H_s = \frac{2}{5} \lambda_s \frac{q^{\frac{5}{3}}}{(1+q)^{\frac{7}{6}}} \left(\frac{R_p}{a}\right)^{\frac{3}{2}} \quad (13)$$

## A.3 Total angular momentum

Thus the general expression for the total angular momentum finally is

$$H = H_o + H_p + H_s = \frac{q}{(1+q)^{\frac{13}{6}}} \sqrt{\frac{a(1-e^2)}{R_p}} + \frac{2}{5} \frac{\lambda_p}{(1+q)^{\frac{5}{3}}} \frac{\omega_p}{\omega_c} + \frac{2}{5} \lambda_s \frac{q^{\frac{5}{3}}}{(1+q)^{\frac{7}{6}}} \left(\frac{R_p}{a}\right)^{\frac{3}{2}} \quad (14)$$

In the case of fully synchronized double systems obeying, for instance, the Roche equations involving a circular orbit and nearly spheroidal components, we have then $\omega_p = \omega_o$, $e=0$ and $\lambda \sim 1$. It stems from the equation (8) that

$$H = \frac{q}{(1+q)^{\frac{13}{6}}} \sqrt{\frac{a}{R_p}} + \frac{2}{5} \frac{1+q^{\frac{5}{3}}}{(1+q)^{\frac{7}{6}}} \left(\frac{R_p}{a}\right)^{\frac{3}{2}} \quad (15)$$

For the peculiar case of equisized components for which we can adopt $q \sim 1$

$$H \sim \frac{2^{\frac{5}{6}}}{5} \left(\frac{R_p}{a}\right)^{\frac{3}{2}} + 2^{-\frac{13}{6}} \sqrt{\frac{a}{R_p}} \quad (16)$$

These handy formulae give the angular momentum of any binary systems. The equation (16) indicates that the orbital angular momentum dominates over the spin for any pair of orbiting spheroids.

# References

Behrend, R., Bernasconi, L., Riy, R., Klotz, A., Colas, F., Antonini, P., Augustesen, K., Barbotin, E., Berger, N., Berrouachdi, H., Brochard, E., Cazenave, A., Cavadore, C., Coloma, J., Deconihout, S., Demeautis, C., Dorseuil, J., Dubos, G., Durkee, R., Frappa, E., Hormuth, F., Itkonen, T., Jacques, C., Kurtze, L., Lavayssière, M., Lecacheurx, J., Leroy, A., Manzini, F., Masi, G., Matter, D., Michelsen, R., Nomen, J., Oksanen, A., Pääkkönen, P., Peyrot, A., Pimentel, E., Pray, D., Rinner, C., Sanchez, S., Sonnenberg, K., Sposetti, S., Starkey, D., Stoss, R., Teng, J.-P., Vignand, M., and Waelchli, N. 2006,. Four new binary minor planets: (854) Frostia, (1089) Tama, (1313) Berna, (4492) Debussy. Astron. Astrophys. 446, 1177-1184.

Belton, M.J.S., Chapman, C.R., Thomas, P.C., Davies, M.E., Greenberg, R., Klaasen, K., Byrnes, D., D'Amario, L., Synnott, S., Merline, W., Petit, J.-M., Storrs, A., and B. Zellner, 1995. Bulk density of asteroid 243 Ida from the orbit of its satellite Dactyl. Nature 374, 785-788.

Cartan, H., 1928. Sur la stabilité ordinaire des ellipsoïdes de Jacobi. Proc. International. Math. Congress, Toronto, 1924 (University of Toronto Press), 2-17.

Chandrasekhar, S. 1969. Ellipsoidal figures of equilibrium. Yale Univ. Press, New Haven.

Chapman, C.R., 1978. Asteroid collisions, craters, regolith, and lifetimes. In Asteroids: An Exploration Assesment (D. Morrison and W.C. Wells, eds.), pp. 145-160. NASA Conf. Publ. 2053.

Cook, A.F., 1971. 624 Hektor: a binary asteroid ? Physical studies of minor planets.

Cuk, M., 2007. Formation and Destruction of Small Binary Asteroids, ApJ 659, 1, L57-L60

Darwin, G.H., 1887. On figures of equilibrium of rotating masses of fluid. Phil. Trans. 379-430.

Davis, D.R., Chapman, C.R., Greenberg, R., Weidenschilling, S.J., Harris, A., 1979. Collisional evolution of asteroids: populations, rotations, and velocities. In Gehrels, T. (Ed.), Asteroids. Univ. of Arizona Press, Tucson, pp. 528-557.

Descamps, P., Marchis, F., Michalowski, T., Vachier, F., Colas, F., Berthier, J., Assafin, M.., Dunckel, P.B., Polinska, M., Pych, W., Hestroffer, D., Miller, K., Vieira-Martins, R., Birlan, M., Teng-Chuen-Yu, J.-P., Peyrot, A., Payet, B., Dorseuil, J., Léonie, Y., and T. Dijoux, 2007a. Figure of the double asteroid 90 Antiope from AO and lightcurves observations. Icarus, 187, 482-499.

Descamps, P., Marchis, F., Michalowski, T., Colas, F., Berthier, J., Vachier, F., Teng-Chuen-Yu, J.P., Peyrot, A., Payet, B., Dorseuil, J., Léonie, Y., Dijoux, T., .Berrouachdi, C. Chion Hock, F. Benard, 2007b. Nature of the small main belt asteroid 3169 Ostro. Icarus, 189, 362-369.

Descamps, P., 2007c. Roche figures of doubly synchronous asteroids. To be appeared in Planetary and Space Science.

Dunlap, J.L., and T. Gehrels, 1969. Minor planets. III. Lightcurves of a Trojan asteroid. The Astron. Journal 74, 796-803.

Durda, D., D., and P. Geissler, 1996. The formation of asteroidal satellites in large cratering collisions. BAAS, 28, 1101.

Eriguchi, Y, Hachisu, I., and D. Sugimoto, 1982. Dumb-bell shape equilibria and mass-shedding pear-shape of selfgravitating incompressible fluid. Progress of theoretical physics, 67, 1068-1075.

Farinella, P., Paolicchi, P., Tedesco, E.F., and V. Zappala, 1981. Triaxial equilibrium ellipsoids among the asteroids ?. Icarus 46, 114-123.

Hachisu, I., and Y. Eriguchi, 1984. Binary fluid star. Publ. Astron. Soc. Japan, 36, 259-276.

Harris, W.A., 1996. The rotation rates of very small asteroids: evidence for "rubble pile" structure. Lunar and Planetary Science, XXVII.

Hartmann, W.K., Cruikshank, D.P., 1978.The nature of Trojan asteroid 624 Hektor. Icarus 36, 353-366.

Hestroffer, D., Berthier, J., Descamps, P., Tanga, P., Cellino, A., Lattanzi, M., di Martino, M., and V. Zappalà, 2002. Asteroid (216) Kleopatra. Tests of the radar-derived shape model. Astronomy and Astrophysics, 396, 301-305.

Hestroffer, D., 2004. On Equilibrium Shapes among Binary Asteroids. BAAS 36, 861.

Hestroffer, D., and P. Tanga, 2005. Figures of Equilibrium among Binary Asteroids. BAAS, 37, 1562

Holsapple, K.A., 2001. Equilibrium configurations of solid cohesionless bodies. Icarus, 154, 432-448.

Holsapple, K.A, 2004. Equilibrium figures of spinning bodies with self-gravity. Icarus 172, 272-303.

Howell, E.S., Rivkin, A.S., Nolan, M.C., Margot, J.L., Black, G., Bus, S.J., Hicks, M., Reach, W.T., Jarrett, T.H., Binzel, R.P., 2003. Observations of 2002 NY40: An Ordinary Chondrite? Physical Properties and Morphology of Small Solar System Bodies, 25th meeting of the IAU, Joint Discussion 19, 23 July 2003, Sydney,

Kaasalainen, M., Torpa, J., and K. Muinonen, 2001. Optimization Methods for Asteroid Lightcurve Inversion. II. The Complete Inverse Problem. Icarus 153, 37-51.

Kaasalainen, M., Torpa, J., and J. Piironen, 2002. Models of Twenty Asteroids from Photometric Data. Icarus, 159, 369-395.

Kryszczynska, A., Kwiatowski, T., Hirsh, R., Polinska, M., Kaminski, K. and K. Marciniak, 2005. (809) Lundia. CBET 239, Edt. By Green, D.W.E.

Manzini, F., Behrend, R., Klotz, A., Ostro, S., Benner, L., Giorgini, J., Nolan, M., Hine, A., Margot, J.-L., Magri, C., Shepard, M., Roy, R., Colas, F., Antonini, P., Paakkonen, P., Bernasconi, L., Leroy, A., Oksanen, A., Crippa, R., Poncy, R., Charbonnel, S., Starkey, D., Coloma, J., Cavadore, C., Barbotin, E., Liapasset, J.-M., Farroni, G., and R. Koff, 2006. (1139) Atami. CBET 430, Edt. By Green, D.W.E.

Marchis, F., Hestroffer, D., Cellino, A., Tanga, P., and V. Zappala, 1999. (216) Kleopatra. IAU Circ. 7308, Edt. By Green, D.W.E.

Marchis, F., Descamps, P., Hestroffer, D., Berthier, J., Vachier, F., Boccaletti, A., de Pater, I., Gavel, D., 2003, A three-dimensional solution for the orbit of the asteroidal satellite of 22 Kalliope. Icarus, 165, 112-120.

Marchis, F., Descamps, P., Hestroffer, D., Berthier, J., 2005a. Discovery of the triple asteroidal system 87 Sylvia. Nature, 436, 822-824.

Marchis, F., Hestroffer, D., Descamps, P., Berthier, J., Laver, C., de Pater, I., 2005b. Mass and density of asteroid 121 Hermione from an analysis of its companion orbit. Icarus, 178, 450-464.

Marchis, F., Hestroffer, D., Descamps, P., Berthier, J., Bouchez, A. H., Campbell, R. D., Chin, J. C. Y., van Dam, M. A., Hartman, S. K., Johansson, E. M., Lafon, R. E., Le Mignant, D., de Pater, Imke, Stomski, P. J., Summers, D. M., Vachier, F., Wizinovich, P. L., Wong, M. H., 2006a. A low density of 0.8gcm-3 for the Trojan binary asteroid 617 Patroclus. Nature, 439, 565-567.

Marchis, F., Baek, M., Berthier, J., Descamps, P., Hestroffer, D., Kaasalainen, M., Vachier, F., 2006c. Large adaptive optics of asteroids (LAOSA): Size, shape, and occasionally density via multiplicity. Workshop on Spacecraft Reconnaissance of Asteroid and Comet Interiors (2006), Abstract #3042.

Marchis, F., Wong, M.H., Berthier, J., Descamps, P., Hestroffer, D., Le Mignant, D., and I. de Pater, 2006b. S/2006 (624) 1. IAU Circ., 8732, Edt. by Green, D. W. E.

Marchis, F., Kaasalainen, M., Hom, E.F.Y., Berthier, J., Enriquez, J., Hestroffer, D., Le Mignant, D., and I. de Pater, 2006d. Shape, size and multiplicity of main-belt asteroids. Icarus, 185, 39-63.

Marchis, F., Baek, M., Berthier, J., Descamps, P., Hestroffer, Vachier, F., 2007a. (617) Patroclus. CBET 836, Edt. By Green, D.W.E.

Marchis, F., Descamps, P., Vachier, F., Berthier, J., Hestroffer, D., Baek, M., Harris, A., and D. Nesvorny, 2007b. Main-belt binary asteroidal systems with eccentric orbits. In preparation.

Marchis, F., Baek, M., Descamps, P., Berthier, J., Hestroffer, D., and F. Vachier, 2007c. S/2004 (45) 1. IAU Circ. 8817, Edt. by Green, D. W. E.

Margot, J.L., Nolan, M.C., Negron, V., Hine, A.A., Campbell, D.B., and E.S. Howell, 2003. 1937 UB (Hermes). IAU Circ. 8227, Edt. by Green, D. W. E.

Merline , W.J., Close, L.M., Dumas, C., et al., 2000. BAAS, 32, 1306.

Merline, W. J., Close, L. M., Siegler, N., et al., 2001, IAU Circ., 7741

Merline, W.J., Weidenshilling, S.J., Durda, D., Margot, J.L., Pravec, P., and A.D. Storrs, 2002. Asteroids do have satellites. Asteroids III W.F., Bottke Jr., A. Cellino, P. Paolocchi, and R.P. Binzels eds, University of Arizona Press, Tucson, p.289-312.

Michalowski, T., Kaasalainen, M., Polinska, M., Marciniak, A., Kwiatkowski, T., Kryszczynska, A., and F.P. Velichko, 2006. Photometry and models of selected main belt asteroids. III. 283 Emma, 665 Sabine, and 690 Wratislavia. Astron. Astrophys., 459, 663-668.

Noll, K., 2006. Solar System binaries. Asteroids, Comets, Meteors, Proceedings of the 229th Symposium of the International Astronomical Union held in Búzios, Rio de Janeiro, Brasil Brazil August 7-12, 2005, Edited by Daniela, L., Sylvio Ferraz, M., Angel, F. Julio Cambridge: Cambridge University Press, 301-318

Ostro, S.P., Hudson, R.S., Nolan, M.C., Margot, J.-L., Scheeres, D.J., Campbell, D.B., Magri, C., Giorgini, J.D., and D.K. Yeomans, 2000. Radar Observations of Asteroid 216 Kleopatra. Science, 288, 836-839

Petit, J.-M., Durda, D.D., Greenberg, R., Hurford, T.A., and P.E. Geissler, 1997. The long-term dynamics of Dactyl's orbit. Icarus, 130, 177-197.

Polishook, D., and N. Brosch, 2006. Many binaries among NEAs. NASA workshop "Near-Earth Object Detection, Characterization, and Threat Mitigation" , Colorado, June 2006

Pravec, P., and G. Hahn, 1997. Two-period lightcure of 1994 AW1: Indication of a binary asteroid ? Icarus, 127, 431-440.

Pravec, P., Wolf, M., and L. Šarounová, 1998. Occultation/eclipse events in binary asteroid 1991 VH. Icarus, 133, 79-88.

Pravec, P, and A.W. Harris 2000. Fast and slow rotation of asteroids. Icarus 148, 12-20.

Pravec, P., Kusnirak, P, Warner, B., Behrend, R., Harris, A.W., Oksanen, A., Higgins, D., Roy, R., Rinner, C., Demeautis, C., Van de Abbeel, F., Klotz, A., Waelchli, N., Bagnoud, F.-X., Al-derweireldt, T., Cotrez, V., and L. Brunetto 2003. 1937 UB (Hermes). IAU Circ. 8233.

Pravec, P., Scheirich, P., Kušnirák, P., Šarounová, L., Mottola, S., Hahn, G., Brown, P.,

Esquerdo, G., Kaiser, N., Krzeminski, Z., Pray, D. P., Warner, B. D., Harris, A. W., Nolan, M. C., Howell, E. S., Benner, L. A. M., Margot, J.-L., Galád, A., Holliday, W., Hicks, M. D., Krugly, Yu. N., Tholen, D., Whiteley, R., Marchis, F., Degraff, D. R., Grauer, A., Larson, S., Velichko, F. P., Cooney, W. R., Stephens, R., Zhu, J., Kirsch, K., Dyvig, R., Snyder, L., Reddy, V., Moore, S., Gajdoš, Š., Világi, J., Masi, G., Higgins, D., Funkhouser, G., Knight, B., Slivan, S., Behrend, R., Grenon, M., Burki, G., Roy, R., Demeautis, C., Matter, D., Waelchli, N., Revaz, Y., Klotz, A., Rieugné, M., Thierry, P., Cotrez, V., Brunetto, L., Kober, G., 2006. Photometric survey of binary near-Earth asteroids. Icarus, 181, 63-93.

Richardson, D.C., and Walsh, K.J., 2006, Binary Minor Planets, Ann. Rev Planet. Sci., 34, 47-81.

Ryan, W. H., Ryan, E. V. and C. T. Martinez, 2004. 3782 Celle: Discovery of a Binary System within the Vesta Family of Asteroids. Planetary and Space Science, 52, 1093-1101.

Sheppard, S.S., and D.C. Jewitt, 2004. Extreme Kuiper belt object 2001 $QG_{298}$ and the fraction of contact binaries. The Astron. Journal, 127, 3023-3033.

Storrs, A.D., Dunne, C., Conan, J.-M., and L. Mugnier, 2000. HST imaging observations of asteroid 216 Kleopatra. B.A.A.S, 32, 1487.

Strahler, A.N., 1971. The Earth Sciences, Harper and Row, New York.

Tanga, P., Hestroffer, D., Berthier, J., Cellino, A., Lattanzi, M.G., di Martino, M., and V. Zappalà, 2001. NOTE: HST/FGS Observations of the asteroid (216) Kleopatra. Icarus, 153, 451-454.

Thomas, P.C., Belton, M.J.S., Carcich, B., Chapman, C.R., Davies, M.E., Sullivan, R., and J. Veverka, 1996. The shape of Ida. Icarus, 120, 20-32.

Walsh, K. J. and Richardson, D.C., 2006. Binary near-earth asteroid formation: rubble pile model of tidal disruptions, Icarus, 180, 1, 201-216.

Weidenschilling, S.J., 1980. Hektor: Nature and origin of a binary asteroid. Icarus, 44, 807-809.

Weidenschilling, S.J., 1981. How fast can an asteroids spin? Icarus, 46, 124-126.

Weidenschilling, S.J., Paolicchi, P.Zappala, V. 1989. Do asteroids have satellites? In Asteroids II; Proceedings of the Conference, Tucson, AZ, Mar. 8-11, 1988 (A90-27001 10-91). Tucson, AZ, University of Arizona Press, 1989, p. 643-658.

Weidenschilling, S.J., Marzari, F., Davis, D.R., and C. Neese, 2001. Origin of the double asteroid 90 Antiope: a continuing puzzle. Lunar and Planetary Science XXXII.

**Table 1**: Orbital and physical properties for the selected binaries of the dataset. *a* is the semi-major axis, $P_{orb}$ the orbital period, $\rho$ the bulk density, $D_p$ and $D_s$ the diameters of the primary and the secondary and $P_{spin}$ the sidereal spin period. See text for explanation of the adimensional quantities $\lambda_p$, $\Omega$, $H$. The references for the orbital and physical parameters are given in the last column. Their uncertainties can be found in the given references.

| Asteroid | a (km) | $P_{orb}$ (days) | $\rho$ (g/cm³) | $D_p$ (km) | $D_s$ (km) | $P_{spin}$ (h) | $\lambda_p$ | $\Omega$ | $H$ | Ref. |
|---|---|---|---|---|---|---|---|---|---|---|
| High size ratio binaries | | | | | | | | | | |
| 22 Kalliope* | 1095 | 3.59 | 2.6 | 181.0 | 38.0 | 4.148 | 1.26[a] | 0.50 | 0.27 | [1] |
| 45 Eugenia* | 1174 | 4.76 | 1.1 | 214.6 | 12.7 | 5.699 | 1.24[a] | 0.55 | 0.28 | [2] |
| 87 Sylvia* | 1356 | 3.67 | 1.3 | 286.0 | 18.0 | 5.184 | 1.27[a] | 0.55 | 0.28 | [3] |
| 107 Camilla* | 1249 | 3.72 | 1.9 | 222.6 | 9.0 | 4.844 | 1.31 | 0.49 | 0.26 | [2] |
| 121 Hermione* | 768 | 2.58 | 1.1 | 209.0 | 18.0 | 5.551 | 1.2 | 0.57 | 0.27 | [5] |
| 130 Elektra* | 1319 | 5.26 | 1.4 | 202.0 | 7.0 | 5.225 | 1.23[e] | 0.53 | 0.26 | [6] |
| 283 Emma* | 581 | 3.36 | 0.8 | 148.1 | 10.0 | 6.888 | 1.24[c] | 0.53 | 0.27 | [6] |
| 379 Huenna* | 3350 | 87.6 | 0.9 | 97.6 | 6.0 | 7.022 | 1.20 | 0.49 | 0.24 | [6] |
| 243 Ida | 108 | 1.54 | 2.4 | 30.45 | 1.4 | 4.6336 | 2.35[d] | 0.44 | 0.42 | [8] |
| Contact-binaries | | | | | | | | | | |
| 216 Kleopatra | 0 | 0 | [3.5] | 135.1 | 3.6 | 5.385 | 3.65[b] | 0.33 | 0.48 | [7] |
| 624 Hektor | 1178 | 3.61 | 2.2 | [225.0] | 15.0 | 6.921 | 3.90[f] | 0.33 | 0.51 | [10] |
| 3169 Ostro | 4.8 | 0.27 | 2.6 | 3.6 | 3.1 | 6.509 | 1.17[f] | 0.31 | 0.46 | [12] |
| Synchronous double asteroids | | | | | | | | | | |
| 69230 Hermes | 1.2 | 0.58 | 1.9 | 0.63 | 0.56 | 13.920 | 1.00[g] | 0.17 | 0.48 | [13] |
| 90 Antiope* | 171 | 0.68 | 1.2 | 87.9 | 83.8 | 16.505 | 1.05[f,g] | 0.18 | 0.49 | [4] |
| 1313 Berna | 24 | 1.06 | 1.2 | 15.2 | 15.2 | 25.464 | 1.20[g] | 0.12 | 0.51 | [11] |
| 4492 Debussy | 18 | 1.11 | 0.9 | 8.1 | 8.1 | 26.606 | 1.20[g] | 0.13 | 0.51 | [11] |
| 1089 Tama | 22 | 0.68 | 2.5 | 9.2 | 9.2 | 16.440 | 1.20[g] | 0.13 | 0.53 | [11] |
| 854 Frostia | 34 | 1.57 | 0.9 | 11.7 | 11.7 | 37.728 | 1.20[g] | 0.09 | 0.57 | [11] |
| 617 Patroclus* | 645 | 4.28 | 0.8 | 121.8 | 112.6 | 101.00 | 1.00[f,g] | 0.04 | 0.74 | [9] |

*: LAOSA Objects
[a]: Shape model from Kaasalainen et al. (2002)
[b]: Shape model from Ostro et al. (2000)
[c]: Shape model from Michalowski et al. (2006)
[d]: Shape model from Thomas et al. (1996)
[e]: Shape model from Marchis et al. (2006d)
[f]: Shape model from Roche ellipsoids
[g]: For the equisized binary systems, $\lambda_p = \lambda_s$

[1] Marchis et al. (2003) ; [2] Marchis et al., in preparation; [3] Marchis et al. (2005a) ; [4] Descamps et al. (2007a); [5] Marchis et al. (2005b); [6] Marchis et al. (2007a) ; [7] Ostro et al. (2000) ; [8] Belton et al. (1995); [9] Marchis et al. (2006a) ; [10] Marchis et al. (2006b) ; [11] Berhend et al. (2006) ; [12] Descamps et al. (2007b); [13] Ostro et al. (2003)

**Figure 1:** High size ratio, contact, and synchronous binaries plotted with respect to scaled spin rate and specific angular momentum. Equilibrium sequences of self-gravitating, rotating fluid masses (Maclaurin, Jacobi, Dumb-bell, Darwin and Roche) have been superimposed to the dataset. Three main isolated regions of binary systems of different nature are highlighted. Roche sequences have been constructed for three values of the mass ratio $s$ (0.5, 0.8, 1.0) and they terminate at their Roche limit. In the case of congruent ellipsoids ($s=1$), Darwin and Roche sequences significantly differ for smaller separations ($\Omega \sim 0.3$) due to the less consistent nature of Roche's problem.

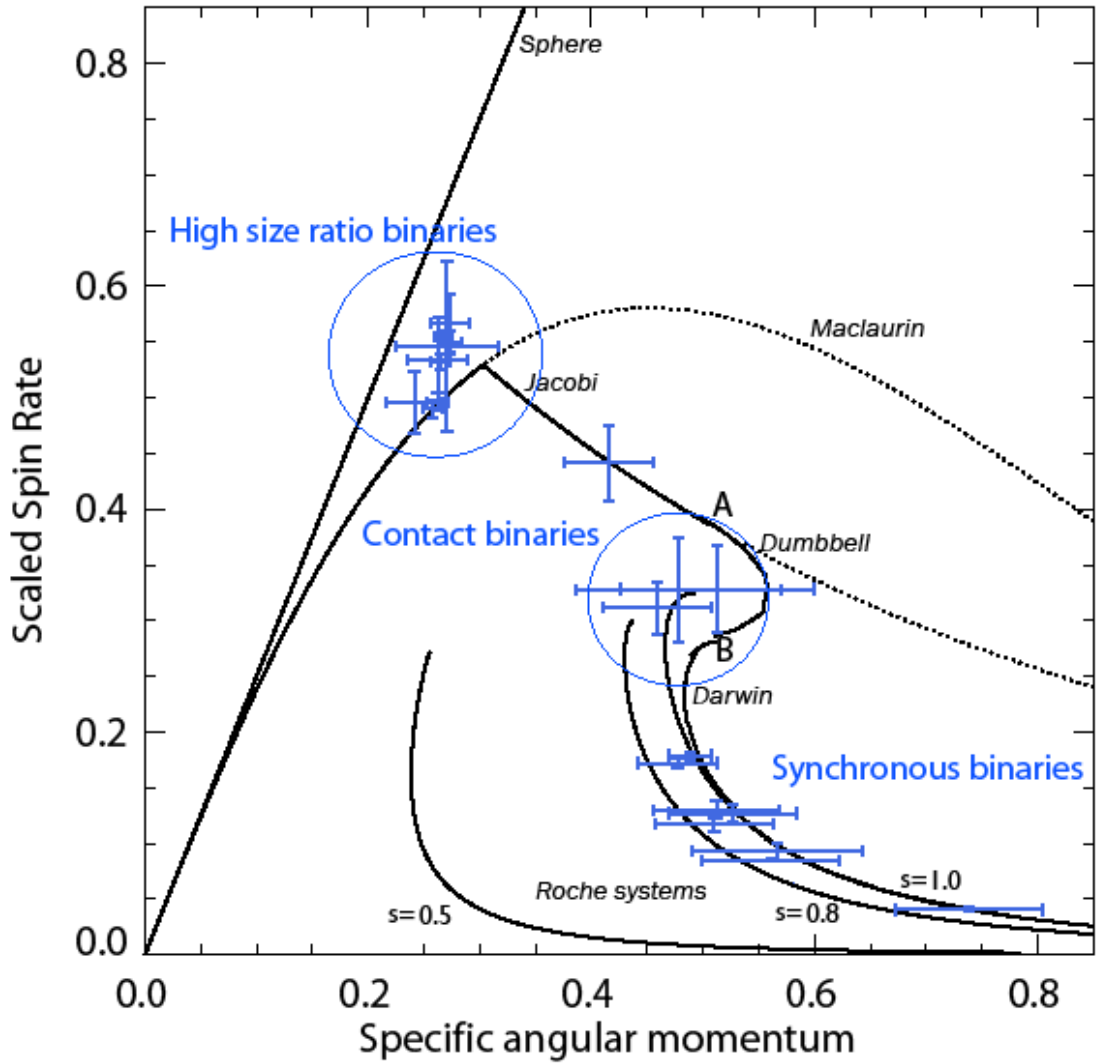

**Figure 2:** Spin curves for an oblate solid with various angles of friction φ (after Holsapple 2004). States on the limit curves are at the maximum spin possible for that angle of friction (solid lines). For each friction angle, there are two curves, one upper limit and one lower limit and all states between are possible equilibrium states. Curves (dotted-lines) of iso-angular momentum are also reported. Binary asteroids with a high primary-to-secondary size ratio have been superimposed from the knowledge of their scaled spin rate and angular momentum given in Table 1. They are all located very close to the stability limit (bold line) along a curve of roughly similar angular momentum. The intersection of the stability limit curve with the iso-angular momentum curve corresponding to H=0.265 gives the starting point of instability which implies a minimum value of the friction angle of φ ∼11°.

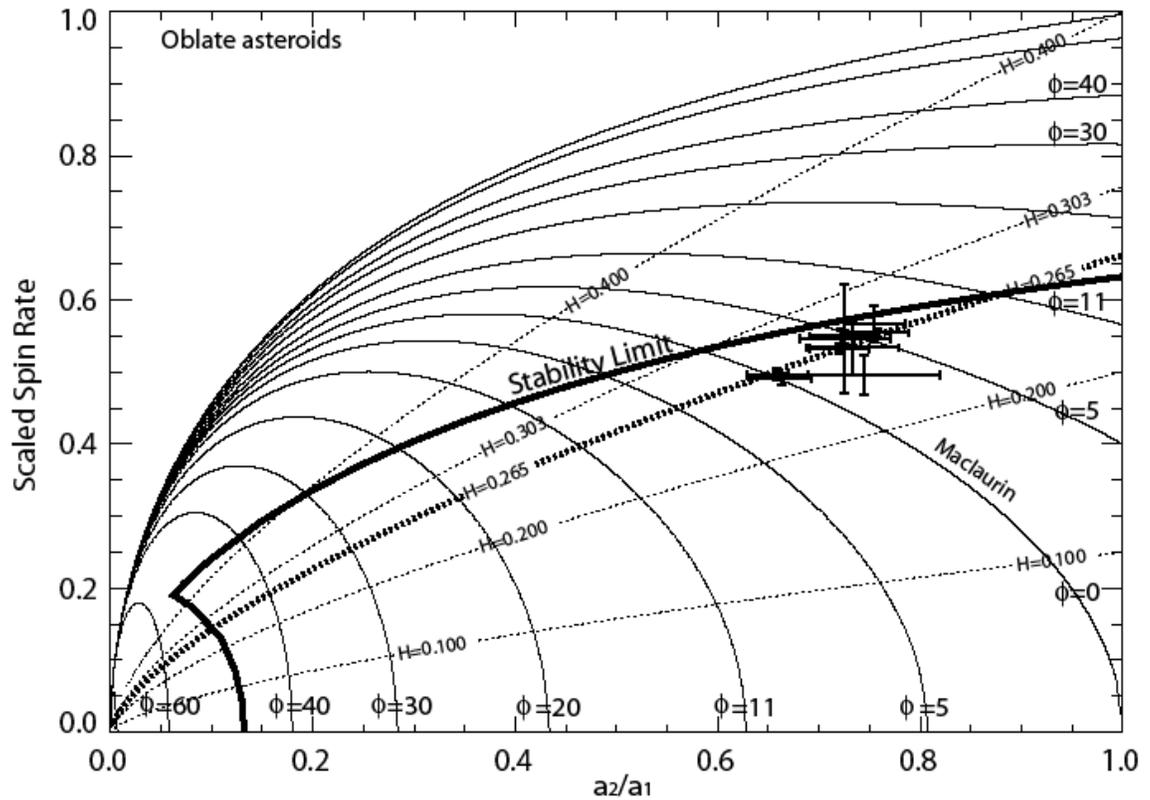

**Figure 3:** Spin curves for a prolate solid with various angles of friction ϕ (after Holsapple 2004). States on the limit curves are at the maximum spin possible for that angle of friction (solid lines). For each friction angle, there are two curves, one upper limit and one lower limit and all states between are possible equilibrium states. Curves (dotted-lines) of iso-angular momentum are also reported. Binary asteroids with a high primary-to-secondary size ratio have been superimposed from the knowledge of their scaled spin rate and angular momentum given in Table 1. They are all located very close to the stability limit (bold line) along a curve of roughly similar angular momentum. The intersection of the stability limit curve with the iso-angular momentum curve corresponding to H=0.265 gives the starting point of instability which implies a minimum value of the friction angle of ϕ ~14°.

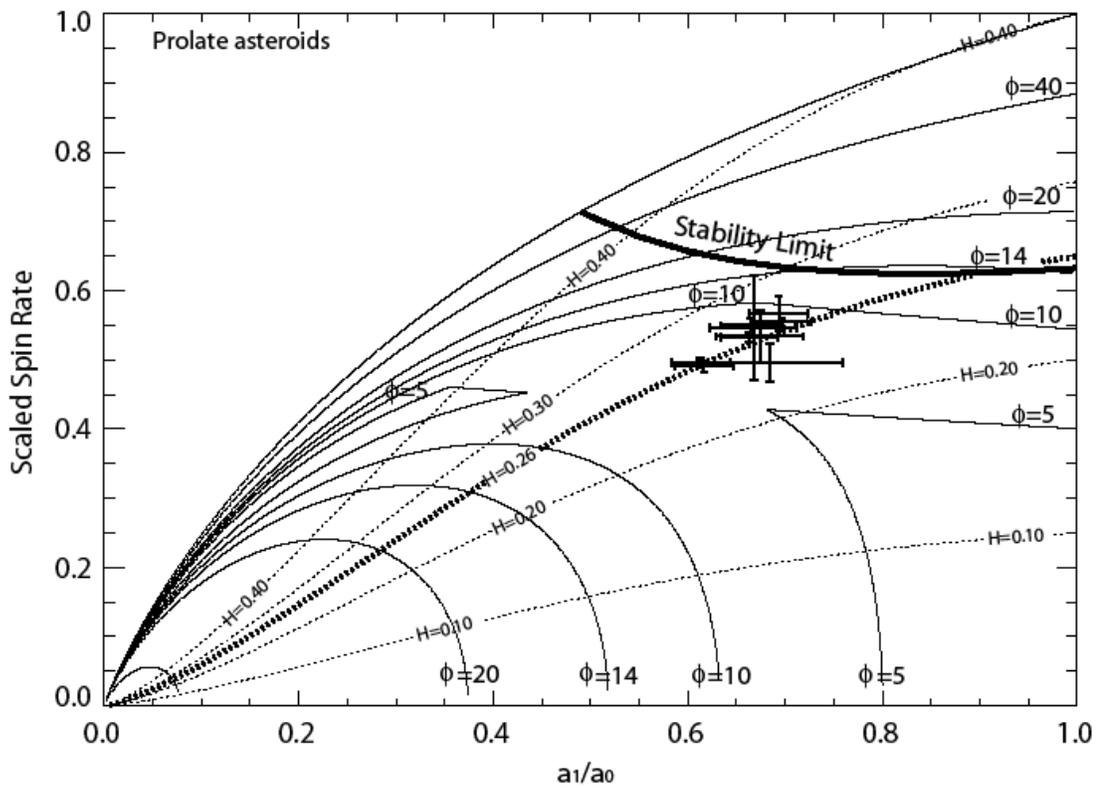